\documentclass[aip,amsmath,amssymb,reprint,jap]{revtex4-2}
\usepackage{graphicx}
\usepackage{epstopdf}
\usepackage{dcolumn}
\usepackage{bm}
\usepackage[utf8]{inputenc}
\usepackage[T1]{fontenc}
\usepackage{mathptmx}
\usepackage{xcolor}
\begin{document}

\title[Fast magnetization reversal of a magnetic nanoparticle induced by cosine chirp microwave field pulse]{Fast magnetization reversal of a magnetic nanoparticle induced by cosine chirp microwave field pulse}

\author{M. T. Islam}
\email{torikul@phy.ku.ac.bd}
\affiliation{Physics Discipline, Khulna University, Khulna 9208, Bangladesh}

\author{M. A. S. Akanda}
\affiliation{Physics Discipline, Khulna University, Khulna 9208, Bangladesh}

\author{M. A. J. Pikul}
\affiliation{Physics Discipline, Khulna University, Khulna 9208, Bangladesh}

\author{X. S. Wang}
\email{justicewxs@hnu.edu.cn}
\affiliation{School of Physics and Electronics, Hunan University, Changsha 410082, China}


\begin{abstract}
We investigate the magnetization reversal of single-domain magnetic nanoparticle driven by the circularly polarized cosine chirp microwave pulse (CCMP). The numerical findings, based on the Landau-Lifshitz-Gilbert equation, reveal that the CCMP is by itself capable of driving fast and energy-efficient magnetization reversal. The microwave field amplitude and initial frequency required by a CCMP are much smaller than that of the linear down-chirp microwave pulse. This is achieved as the frequency change of the CCMP closely matches the frequency change of the magnetization precession which leads to an efficient stimulated microwave energy absorption (emission) by (from) the magnetic particle before (after) it crosses over the energy barrier. We further find that the enhancement of easy-plane shape anisotropy significantly reduces the required microwave amplitude and the initial frequency of CCMP. We also find that there is an optimal Gilbert damping for fast magnetization reversal. These findings may provide a pathway to realize the fast and low-cost memory device.
\end{abstract}

\maketitle

\thispagestyle{empty}
\section{\label{sec:introduction}Introduction}
Achieving fast and energy-efficient magnetization reversal of high anisotropy materials has drawn much attention since it has potential application in non-volatile data storage devices \cite{sun2000,woods2001,zitoun2002} and fast data processing \cite{hillebrands2003}. For high thermal stability and  low error rate, high anisotropy materials are required \cite{mangin2006} in device application. But one of the challenging issues is to find out the way which can induce the fastest magnetization reversal with minimal energy consumption. Over the last two decades, many magnetization reversal methods has been investigated, such as by constant magnetic fields \cite{hubert1998,sun2005}, by the microwave field of constant frequency, either with or without a polarized electric current \cite{bertotti2001,sunz2006,zhu2010} and by spin-transfer torque (STT) or spin-orbit torque (SOT)~\cite{slonczewski1996,berger1996,tsoi1998,sun1999,bazaliy1998,Katine2000,Waintal2000,sun2000a,suns2003,stiles2002,bazaliys2004,koch2004,li2004,wetzels2006,manchon2008,miron2010,miron2011,liu2012}. However, all the means are suffering from their own limitations. For instance, in the case of external magnetic field, reversal time is longer and has scalability and field localization issues \cite{hubert1998}. In case of the constant microwave field driven magnetization reversal, the large field amplitude and the long reversal time are emerged as limitations \cite{denisov2006,okamoto2008,tanaka2013}. In the case of the STT-MRAM, the threshold current density is a large and thus,  Joule heat which may lead the device malfunction durability and reliability issues \cite{grollier2003,morise2005,taniguchi2008,SUZUKI200993,zsun2006,wang2007,wange2008}. Moreover, there are several studies showing magnetization reversal induced by microwaves of time-dependent frequency \cite{thirion2003,rivkin2006,wangc2009,barros2011,barrose2013,klughertzx2014,islam2018}. In the study \cite{thirion2003}, the magnetization reversal, with the assistance of external field, is obtained by a radio-frequency microwave field pulse. Here, a dc external field acts as the main reversal force. In the study \cite{rivkin2006}, to obtain magnetization reversal,  the applied microwave frequency needs to be the same as the resonance frequency, and in the studies \cite{barros2011,barrose2013}, optimal microwave forms are constructed. These microwave forms are difficult to be realized in practice. The study \cite{klughertzx2014} reports magnetization reversal induced by the microwave pulse, but the pulse is applied such that the magnetization just crosses over the energy barrier, i.e., only positive frequency range ($+f_0$ to 0) is employed.

A recent study \cite{islam2018} has demonstrated that the circularly polarized linear down-chirp microwave pulse (DCMP) (whose frequency linearly decreases with time from the initial frequency $+f_0$ to $-f_0$) can drive fast magnetization reversal of uniaxial nanoparticles. The working principle of the above model is that the DCMP triggers stimulated microwave energy absorption (emission) by (from) the magnetization before (after) crossing the energy barrier. However, the efficiency of triggered microwave energy absorption or emission depends on how closely the frequency of chirp microwave pulse matches the magnetization precession frequency. In DCMP-driven case, the frequency linearly decreases from $f_0$ to $-f_0$ with time but, in fact, the decrement of magnetization precession frequency is not linear \cite{cai2013r, cai2010i} during magnetization reversal . So the frequency of DCMP only roughly matches the magnetization precession frequency. Thus, the DCMP triggers inefficient energy absorption or emission and the required microwave amplitude is still large.

Therefore, to achieve more efficient magnetization reversal, we need to find a microwave pulse of proper time-dependent frequency that matches the intrinsic magnetization precession frequency better. In this study, we demonstrate that a cosine chirp microwave pulse (CCMP), defined as a microwave pulse whose frequency sweeps in a cosine function with time
from $+f_0$ to $-f_0$ in first half-period of the microwave pulse, is capable of driving the fast and energy efficient magnetization reversal. This is because the frequency change of the CCMP matches the nonlinear frequency change of magnetization precession better than the DCMP. In addition, this study emphasizes how the shape anisotropy influences the required parameters of CCMP and how the Gilbert damping affects the magnetization reversal. We find that the increase of easy-plane shape anisotropy makes the magnetization reversal easier. The materials with larger damping are better for fast magnetization reversal. These investigations might be useful in device applications.

\section{\label{sec:model}Analytical Model and Method}
\begin{figure}
    \centering
	\includegraphics[width=85mm]{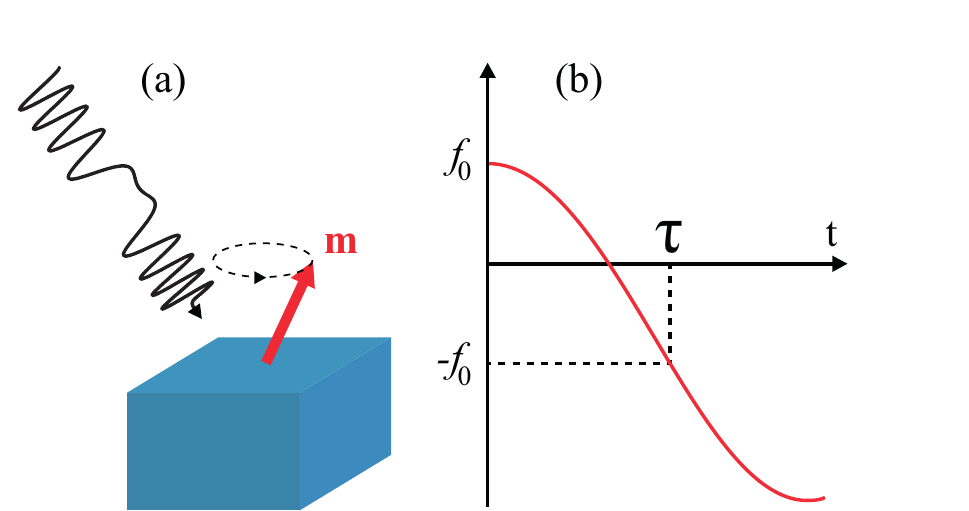}
    \caption{\label{Fig1}(a) Schematic diagram of the system in which $\mathbf{m}$ represents a unit vector of the magnetization. A circularly polarized cosine (nonlinear) chirp  microwave pulse is applied onto the single domain nanoparticle. (b) The frequency sweeping (from $+f_0$ to $-f_0$) of a cosine chirp  microwave pulse.}
\end{figure}

We consider a square magnetic nanoparticle of area $S$ and thickness $d$ whose uniaxial easy-axis anisotropy directed in the $z$-axis as shown in FIG. \ref{Fig1}(a). The size of the nanoparticle is chosen so that the magnetization is considered as a macrospin represented by the unit vector $\mathbf{m}$ with the magnetic moment $SdM_{s}$, where $M_s$ is the saturation magnetization of the material.
The demagnetization field can be approximated by a easy-plane shape anisotropy. The shape anisotropy field coefficient is $h_\text{shape}=
-\mu_0(N_z-N_x)M_s$, and the shape anisotropy field is $\mathbf{h}_\text{shape}=h_\text{shape}m_z \hat{\mathbf{z}}$,
where $N_z$ and $N_x$ are demagnetization factors \cite{dubowik1996,aharoni1998}and $\mu_0 = 4\pi \times 10^{-7} \text{\:N}/\text{A}^2$ is the vacuum magnetic permeability.
The strong uniaxial magnetocrystalline anisotropy $\mathbf{h}_\text{ani}=h_\text{ani}m_z \hat{\mathbf{z}}$ dominates the total anisotropy so that the magnetization of the nanoparticle has two stable states, i.e., $\mathbf{m}$ parallel to $\hat{\mathbf{z}}$ and $-\hat{\mathbf{z}}$.

The magnetization dynamics $\mathbf{m}$ in the presence of circularly polarized CCMP is governed by the Landau-Lifshitz-Gilbert (LLG) equation \cite{gilbert2004p}

\begin{equation}
\frac{d\mathbf{m}}{d t} = - \gamma \mathbf{m} \times \mathbf{h}_{\text{eff}} + \alpha \mathbf{m} \times \frac{d\mathbf{m}}{d t},
\end{equation}
where  $\alpha$ and $\gamma$ are the dimensionless Gilbert damping constant and the gyromagnetic ratio, respectively, and $\mathbf{h}_\text{eff}$ is the total effective field which includes the microwave magnetic field $\mathbf{h}_\text{mw}$, and the effective anisotropy field $\mathbf{h}_\text{k}$ along $z$ direction (note that although we consider small nanoparticles that can be treated as macrospins approximately, we still perform full micromagnetic simulations with small meshes and full demagnetization field to be more accurate).

The effective anisotropy field can be expressed in terms of uniaxial anisotropy $\mathbf{h}_\text{ani}$ and shape anisotropy $\mathbf{h}_\text{shape}$ as $\mathbf{h}_\text{k} = \mathbf{h}_\text{ani} + \mathbf{h}_\text{shape} = \left[h_\text{ani} - \mu_0 (N_z-N_x)M_\text{s}\right] m_z \hat{\mathbf{z}}$.  Thus, the resonant frequency of the nanoparticle is obtained from the well-known Kittel formula
\begin{align} f_0=\frac{\gamma}{2\pi}\left[h_\text{ani} - \mu_0 (N_z-N_x)M_\text{s}\right].
  \label{reso}
\end{align}.

For microwave field-driven magnetization reversal from the LLG equation, the rate of energy change is expressed as
\begin{align}
    \frac{dE}{dt} = - \alpha \gamma \left| \mathbf{m} \times \mathbf{h}_\text{eff} \right|^2 - \mathbf{m} \cdot \frac{d{\mathbf{h}}_\text{mw}}{dt}.
    \label{eng}
\end{align}
The first term is always negative since damping $\alpha$ is positive. The second term can be either positive or negative for time-dependent external microwave field. Therefore, the microwave field pulse can trigger the stimulated energy absorption or emission, depending on the angle between the instantaneous magnetization $\mathbf{m}$ and $\frac{d{\mathbf{h}}_\text{mw}}{dt}$.

Initially, because of easy-axis anisotropy, the magnetization prefers to stay in one of the two stable states, $\pm\hat{\mathbf{z}}$, corresponding to two energy minima. The objective of magnetization reversal is to get the magnetization from one stable state to the other. Along the reversal process, the magnetization requires to overcome an energy barrier at $m_z=0$  which separates two stable states. For fast magnetization reversal, the external field is required to supply the necessary energy to the magnetization until crossing the energy barrier and, after crossing the energy barrier, the magnetization releases energy through damping and$\slash$or the external field is required to extract (by negative work done) energy from the magnetization. It is mentioned that there is as intrinsic anisotropy field $\mathbf{h}_\text{ani}$ due to the anisotropy which induces a magnetization natural$\slash$resonant frequency proportional to $m_z$. When magnetization goes from one stable state to another, the magnetization resonant frequency decreases while the magnetization climbs up and becomes zero momentarily while crossing the energy barrier and then increases with the opposite precession direction while it goes down from the barrier. In principle, for fast and energy-efficient reversal, one requires a chirp microwave pulse whose frequency always matches the magnetization precession frequency to ensure the term $\mathbf{m} \cdot \Dot{\mathbf{h}}_\text{mw}$ to be maximal (minimal) before (after) crossing the energy barrier. The study \cite{islam2018}, employs the DCMP (whose frequency linearly decreases with time) to match the magnetization precession frequency roughly. In fact, during the reversal from  $m_z=+1$  to $m_z=-1$ , the decreasing of the resonant frequency ( while the spin
climbs up the energy barrier) and increasing of the resonant frequency (while it goes
down from the barrier) are not linear\cite{cai2013r, cai2010i}.  This leads us to consider a cosine chirp microwave
pulse (CCMP) (a microwave pulse whose frequency decreases non-linearly with time) in order to match the change of magnetization precession frequency closely. Thus the CCMP might trigger more efficient stimulated microwave absorptions
(emissions) by (from) magnetization before (after) the spin
crosses the energy barrier to induce fast and energy-efficient magnetization reversal.

\begin{figure}
    \centering
    \includegraphics[width=85mm]{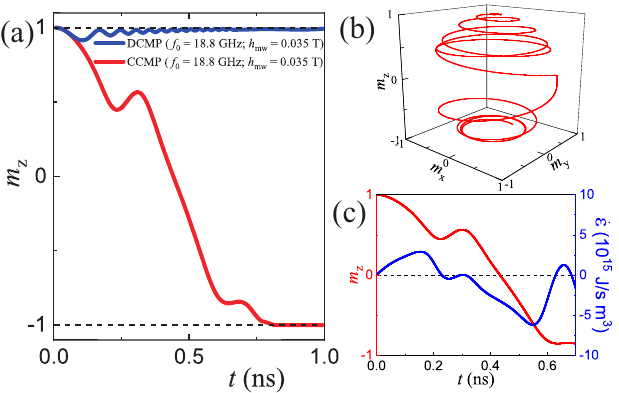}
    \caption{\label{fig:Fig2} Model parameters of nanoparticle of  $M_\text{s}$ = $10^6$ A/m, $H_\text{k}$ = $0.75$ T, $\gamma$ = $1.76\times 10^{11}$ rad/(T$\cdot$s), and $\alpha= 0.01$. (a) Temporal evolutions of $m_z$ of $V = (8 \times 8 \times 8)$ $\text{\:nm}^3$ driven by the CCMP (with the minimal $h_\text{mw} = 0.035 \text{\:T}$, $f_0 = 18.8 \text{\:GHz} $ and  $R$ = 0.32 GHz) (red line) and DCMP (with $h_\text{mw} = 0.035 \text{\:T}$, $f_0 = 18.8 \text{\:GHz}$ and $R$ = 1.53 GHz ) (blue line). For CCMP case, (b) the corresponding magnetization reversal trajectories  and (c) temporal evolutions of $m_z$ (red lines) and the energy changing rate $\dot{ \varepsilon }$ of the magnetization against time (blue lines).}
\end{figure}

In order to substantiate the above mentioned prediction, we apply  a circularly polarized cosine down-chirp  microwave pulse in the $xy$ plane of the nanoparticle and solve the LLG equation numerically using the MUMAX3 package \cite{vansteenkiste2014}. The cosine chirp microwave pulse (CCMP) takes the form $\mathbf{h}_\text{mw} = h_\text{mw} \left[ \cos\phi(t) \hat{\mathbf{x}} + \sin\phi(t) \hat{\mathbf{y}}\right]$, where $h_\text{mw}$ is the amplitude of the microwave field and $\phi(t)$ is the phase.  Since the phase $\phi(t)$ is $2 \pi f_0 \cos \left(2 \pi R t \right) t$, where $R$ (in  units  of GHz) is the controlling parameter, the instantaneous frequency $f(t)$ of CCMP is obtained as $f(t) =\frac{1}{2\pi} \frac{d\phi}{dt}= f_0 \left[\cos \left(2 \pi Rt \right) - \left(2 \pi Rt\right) \sin \left(2 \pi Rt\right) \right]$ which decreases  with time from $f_0$ to final $-f_0$  at a time dependent chirp rate $\eta(t)$ (in units of $\text{ns}^{-2}$) as shown in FIG. \ref{Fig1}(b).
The chirp rate takes the form $\eta(t) = - f_0 \left[\left({4 \pi} R\right) \sin\left({2 \pi Rt} \right) + \left({2 \pi}R \right)^2 t \cos\left({2 \pi Rt} \right) \right]$.

According to the applied CCMP, the second term of right hand side of Eq. \eqref{eng}, i.e., the energy changing rate can be expressed as
\begin{align}
   \dot{\varepsilon} = - H_{mw} \sin\theta(t) \sin\Phi(t) \left[ \frac{\phi(t)}{t} - \frac{d}{dt} \left(\frac{\phi(t)}{t}\right) t \right]
      \label{eq:Io}
\end{align}
where $\Phi(t)$ is the angle between $\mathbf{m}_t$ (the in-plane component of $\mathbf{m}$ and $\mathbf{h}_\text{mw}$.
Therefore, the microwave field pulse can trigger the stimulated energy absorption (before crossing the energy barrier) with $-\Phi(t)$  and emission (after crossing the energy barrier) with  $\Phi(t)$.

The material parameters of this study are chosen from typical experiments on microwave-driven magnetization reversal as $M_\text{s} = 10^6 \text{\:A}/ \text{m}$, $h_\text{ani} = 0.75 \text{\:T}$, $\gamma = 1.76\times 10^{11} \text{\:rad}/(\text{T}\cdot\text{s})$, exchange constant $A = 13 \times 10^{-12} \text{\:J}/\text{m}$ and $\alpha = 0.01$.  Although, the  strategy and other findings of this study would work for other materials also.
The cell size $(2 \times 2 \times 2) \text{\:nm}^3$ is used in this study. We consider the switching time window $1 \text{\:ns}$ at which the magnetization switches\slash reverses to $m_z = - 0.9$.

\begin{table*}
\caption{\label{tab:table1} Shape anisotropy coefficient, resonant frequency  $f_0$, simulated frequency, $f_0$ and frequency-band}
\begin{ruledtabular}
\begin{tabular}{c c c c c}
Cross$-$sectional area &Shape anisotropy coefficient
&Resonant frequency, & Simulated minimal frequency, & Simulated frequency-band\\$S$ (nm$^2$)&\text{$h_\text{shape}$ (T)}&$f_0$ (GHz)& $f_0$ (GHz)&(GHz)\\
\hline\\
 $S_1$ & 0.09606 & 18.3 & 17.8  &  17.8$-$19.9 \\
 $S_2$ & 0.17718 & 16 & 14.9 & 14.9$-$18.3 \\
 $S_3$ & 0.2465 & 14.1 & 13.7 &13.7$-$15.4 \\
 $S_4$ & 0.3064 & 12.4 & 12.2 &12.2$-$13.8 \\
 $S_5$ & 0.3588 &  11 &10.7  & 10.7$-$12.3 \\
 $S_6$ & 0.4049 & 9.6 &  8.8 & 8.8$-$11.5 \\
 $S_7$ & 0.4459 & 8.5 & 7.7 & 7.4$-$8.7 \\
\end{tabular}
\end{ruledtabular}
\end{table*}
\section{\label{sec:numeric}Numerical Results}

We first investigate the possibility of reversing the magnetization of cubic sample $(8 \times 8 \times 8) \text{\:nm}^3$ by the cosine chirp microwave pulse (CCMP). Accordingly, we apply the CCMP with the microwave amplitude $h_\text{mw} = 0.045 \text{\:T}$, initial frequency $f_0=21 \text{\:GHz}$ and  $R = 1.6$ $ \text{ns}^{-1}$ which are same as estimated in the study \cite{islam2018}), to the sample and found that CCMP can drive the fast magnetization reversal.  Then, we search the minimal $h_\text{mw}$, $f_0$, and $R$ of the CCMP such that the fast reversal is still valid. Interestingly, the CCMP with significantly smaller parameters i.e., $h_\text{mw}=0.035$ T, $f_0=18.8 \text{\:GHz}$  and optimal $R =0.32$ $\text{ns}^{-1}$, is capable of reversing the magnetization efficiently shown by red line in FIG. \ref{fig:Fig2}(a).

Then we intend to show how efficient the CCMP driven magnetization reversal compare to DCMP driven case. For fair comparison, we choose the same pulse duration $\tau$ (as shown in Fig. 1, $\tau$ is the time at which the frequency changes from $+f_0$
to $-f_0$). For the CCMP, we solve $\cos(2\pi R \tau)-(2\pi R \tau)\sin(2\pi R \tau)=-1$, which gives the relation $\tau=\frac{1.307}{2\pi R}$. But, in case of DCMP, we know that $\tau=\frac{2f_0}{\eta}$ and for $f_0=18.8 \text{\:GHz}$, the chirp rate becomes $\eta=57.86$ $\text{ns}^{-2}$ and hence find the parameter $R (=1/\tau)$ = 1.53 $\text{ns}^{-1}$. Then we apply the DCMP with the  $h_\text{\:mw}=0.035$ T, $f_0=18.8 \text{\:GHz}$ (which are same as  CCMP-driven case), and $R$ =1.53 $\text{ns}^{-1}$ and found that the magnetization only precesses around the initial state, i.e., the DCMP is not able to reverse the magnetization as shown by blue line in FIG. \ref{fig:Fig2}(a). So, it is mentioned that the CCMP can reverse the magnetization with lower energy consumption which is the desired in device application. To be more explicit, the trajectories of magnetization reversal induced by CCMP is shown FIG. \ref{fig:Fig2}(b) which shows the magnetization reverses swiftly.For further justification of CCMP driven reversal, we calculate the energy changing rate $dE/dt$ refers to \eqref{eq:Io} by determining the angles $\Phi(t)$ and $\theta(t)$ and plotted with time in FIG. \ref{fig:Fig2}(c). The stimulated energy absorption (emission) peaks are obtained before (after) crossing the energy barrier as expected for faster magnetization reversal. This is happened because the frequency of the CCMP closely matches the frequency of magnetization precession frequency, i.e., before crossing the energy barrier,  the $\Phi(t)$ remains around $-90^{\circ}$ and after crossing the energy barrier $\Phi(t)$ around $90^{\circ}$ to maximize the energy absorption and emission respectively.

\begin{figure}
\centering
	\includegraphics[width=85mm]{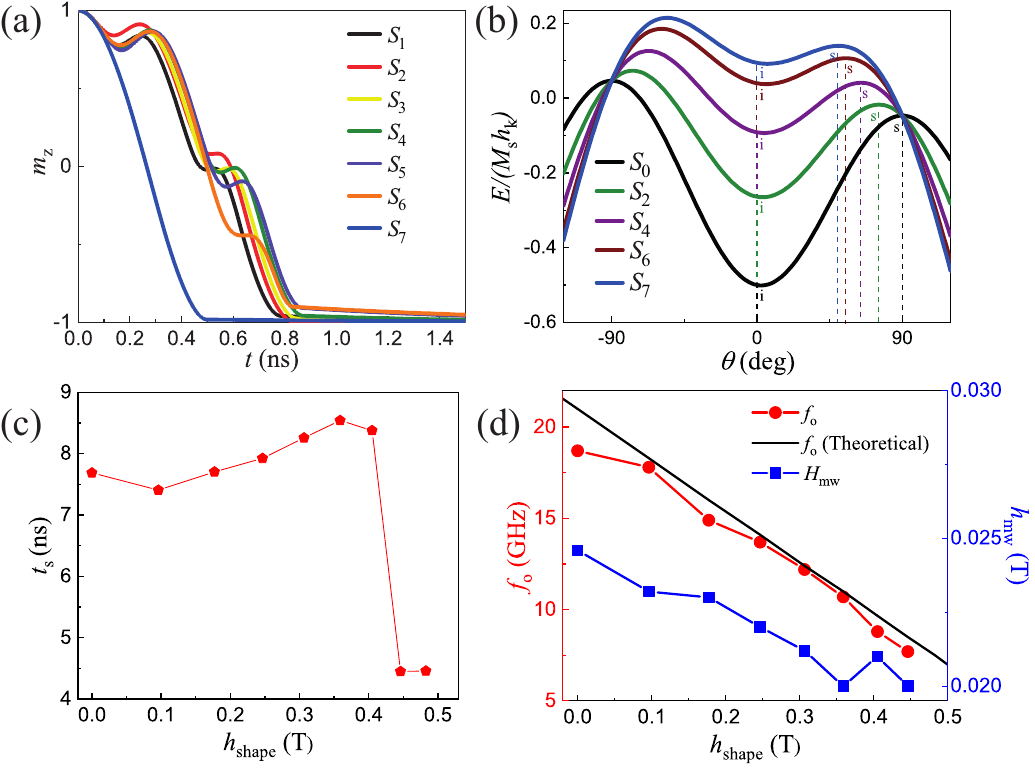}
    \caption{\label{fig:Fig3}(a) Temporal evolution of $m_z$ induced by CCMP (with $h_\text{mw}= 0.035 \text{\:T}$ fixed) for different cross-sectional area, $S$. (b) The energy landscape $E$ along the line $\phi = 0$. The symbols i and s represent the initial state and saddle point. (c) $t_s$ as a function of 
    $h_\text{shape}$. (d) The minimal $f_0$ (red dotted) and $h_\text{mw}$ (blue square) as a function of $h_\text{shape}$  while switching time window $1 \text{\:ns}$.}
\end{figure}

Then, this study  emphasizes  how  shape-anisotropy field  $\mathbf{h}_\text{shape}$ affects  the magnetization switching time, microwave amplitude $h_\text{mw}$ and initial frequency $f_0$ of CCMP. Since demagnetization field or shape-anisotropy field $\mathbf{h}_\text{shape}$ should have significant effect on magnetization reversal process as it opposes the magnetocrystalline anisotropy field  $\mathbf{h}_\text{ani}$  which stabilizes the magnetization along two stable states.
Accordingly,  to induce the $h_\text{shape}$ in the sample, we choose square cuboid shape samples and, to increase the strength of $h_\text{shape}$, the cross-sectional area $S=xy$ is enlarged gradually for the fixed thickness $d\space(=z)\space = 8$ nm. Specifically, we focus on the samples of $S_1=10 \times 10 $, $S_2=12 \times 12 $, $S_3=14 \times 14 $, $S_4=16 \times 16 $, $S_5=18 \times 18 $, $S_6=20 \times 20 $ and $S_7=22 \times 22 $ nm$^2$  with $d = 8$ nm.  For the samples of different $S$, by determining the analytic demagnetization factors $N_z$ and  $N_x$ \cite{dubowik1996,aharoni1998d}, the  shape anisotropy field  $  \mathbf{h}_\text{shape} = \mu_0 (N_z-N_x)M_\text{s} m_{z} \hat{\mathbf{z}}$ are determined.  The  $\mathbf{h}_\text{shape}$ actually opposes the anisotropy field  $\mathbf{h}_\text{ani}$ and hence resonance frequencies  $f_0=\frac{\gamma}{2\pi}\left[h_\text{ani} - \mu_0 (N_z-N_x)M_\text{s}\right]$ decreases as shown in the Table \ref{tab:table1}. Then, for the samples of different $S$, with the fixed $h_\text{mw}= 0.035 \text{\:T}$,  the corresponding minimal $f_0$ and optimal $R$ of CCMP are determined through the study of the magnetization reversal.
The temporal evolutions of $m_z$ for different $S$ are shown in FIG. \ref{fig:Fig3}(a) and found that for $S_7 \geq 22 \times 22 $ nm$^2$, the magnetization smoothly reveres with the shortest time. The switching time $t_s$ as a function of the coefficient $h_\text{shape}$ (corresponding to $S$)  is  plotted in FIG. \ref{fig:Fig3}(c). It is observed that, with the increase of $ {h}_\text{shape}$ or $S$, the $t_s$ shows slightly increasing trend but for $S_7$ $ (22 \times 22) $ nm$^2$ or ${h}_\text{shape}=0.4459$ T ,  $t_s$ drops to  0.43 $\text{\:ns}$. For further increment of  $S$ or ${h}_\text{shape}$, $t_s$ remains constant around  0.43 $\text{\:ns}$ which is close to the theoretical limit (0.4$\text{\:ns}$) refers to the study \cite{wang2007}.

This is because the $\mathbf{h}_\text{shape}$ reduces the effective anisotropy  and thus reduces the height of energy barrier  (energy difference between the initial state and saddle point) which is shown in the FIG. \ref{fig:Fig3}(b).  Therefore, after decreasing certain height of energy barrier, the magnetization reversal becomes fastest even with the same filed amplitude $h_\text{mw}= 0.035$ \text{\:T}.
Due to the reduction of height of the energy barrier with the  $h_\text{shape}$, one can expect that the $h_\text{mw}$ and  $f_0$  should also decrease with the increase of the anisotropy coefficient  $\mathbf{h}_\text{shape}$ and theses findings are presented subsequently.

\begin{figure}
    \centering
    \includegraphics[width=85mm]{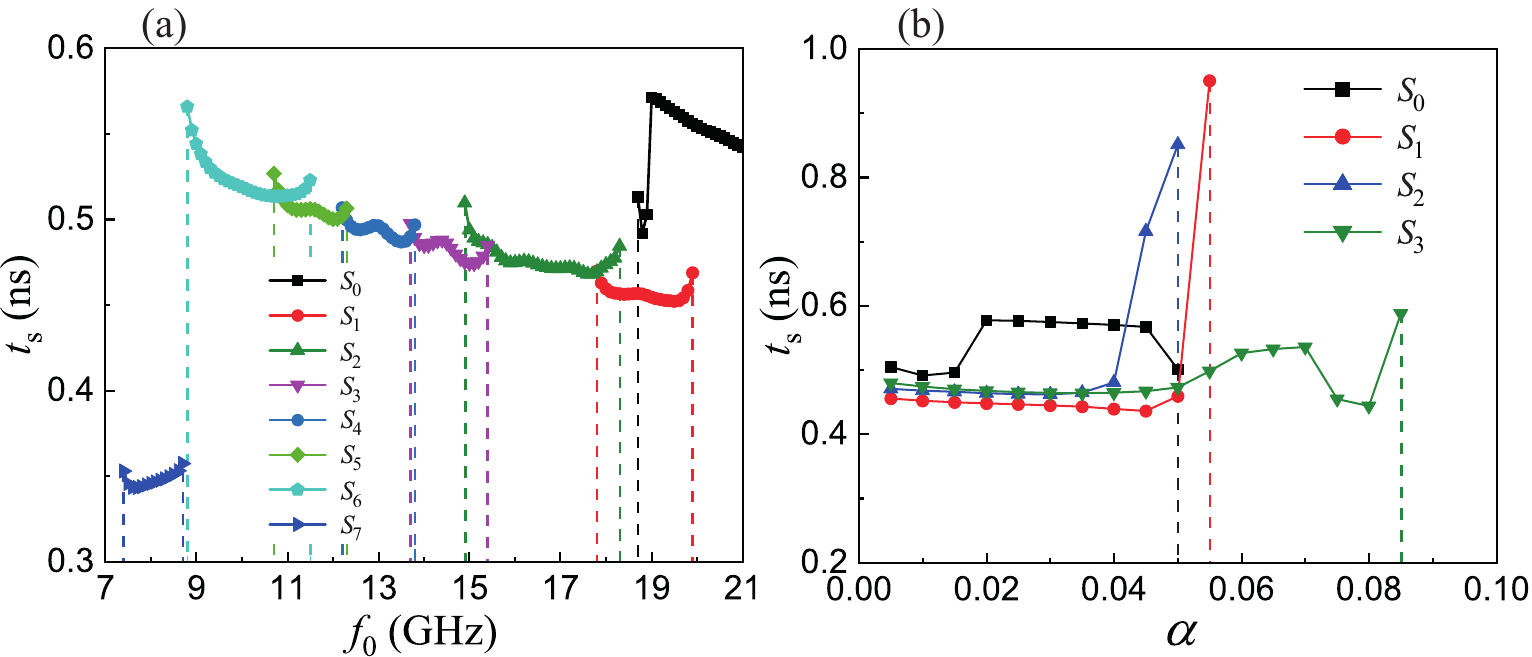}
    \caption{\label{fig:Fig4}(a) Minimal switching time $t_s$ as a function of estimated $f_o$ of CCMP with fixed $h_\text{mw}$  and $R$ corresponding to different $S$. (b) Minimal $t_s$  as a function of Gilbert damping $\alpha$ for different $S$.}
\end{figure}

Here we present the effect of the shape anisotropy $h_\text{shape}$ on the microwave amplitude $h_\text{mw}$ and initial frequency $f_0$  of CCMP.  Purposely, for each sample $S$ or  $h_\text{shape}$, we numerically determine (by tuning $h_\text{mw}$, $f_0$ and optimal $R$) the minimally required  $h_\text{mw}$ and $f_0$ of CCMP for the the magnetization reversal time window 1 ns.  Interestingly, we find that the fast and efficient reversal is valid for a wide range of initial frequency $f_0$. FIG. \ref{fig:Fig4}(a) shows the estimated frequency bands of $f_0$ by vertical dashed lines for different $S$ ( for example, $f_0=19.9 \sim 17.8$ for ${h}_\text{shape}=0.096$ T) for time window $1 \text{\:ns}$. So there is a great flexibility in choosing the initial frequency  $f_0$ which is useful in device application. FIG. \ref{fig:Fig3}(d) shows how the minimal $f_0$ (red circles) and $h_\text{mw}$ (blue square)  decrease with the increase of $h_\text{shape}$. The decay of  $f_0$ is expected  as $h_\text{shape}$ reduces effective anisotropy (refers to Eq.\eqref{reso}). For more justification, in same FIG. \ref{fig:Fig3}(d), theoretically resonant frequency (black solid line) and simulated minimal frequency (red circles) as function of $h_\text{shape}$ indicate the agreement. The minimal frequency $f_0$ always smaller than the theoretical$\slash$resonant frequency. 

Moreover, the decreasing trend of  $h_\text{mw}$ with $h_\text{shape}$ can be attributed by the same reason as the height of the energy barrier decreases with $h_\text{shape}$  (refers to FIG. \ref{fig:Fig3}(b)). For the larger $h_\text{shape}$ or lower height of energy barrier, the smaller microwave field $h_\text{mw}$ can induce the magnetization reversal swiftly.

\begin{figure}
\centering
	\includegraphics[width=85mm]{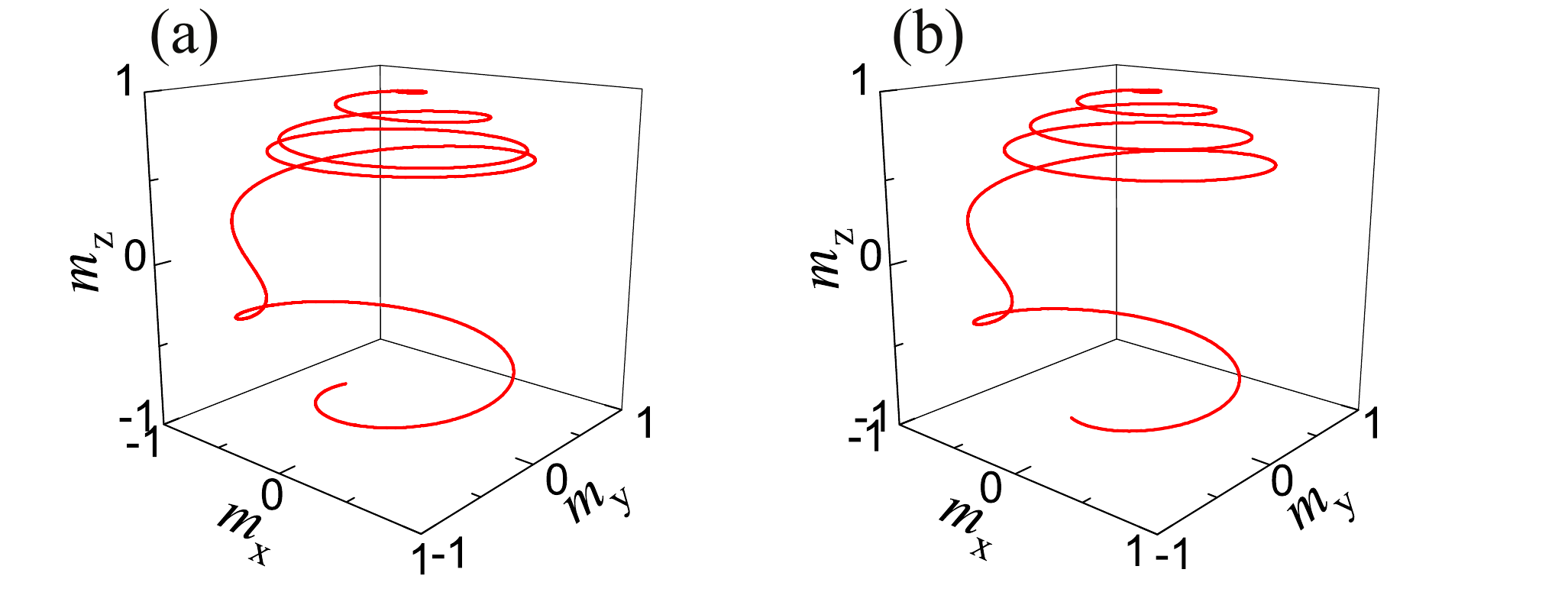}
    \caption{\label{fig:Fig5} Magnetization reversal trajectories of biaxial shape $(10 \times 10 \times 8) \text{\:nm}^3$ driven by CCMP for (a)  $\alpha = 0.010$. (b) $\alpha = 0.045$.}
\end{figure}

The Gilbert damping parameter, $\alpha$ has also a crucial effect on the magnetization magnetization dynamics and hence reversal process and  reversal time \cite{wang2010c,chen2011d,guo2007e}. In case of CCMP-driven magnetization reversal, smaller (larger) $\alpha$ is preferred while the magnetization  climbing the energy barrier (the magnetization goes down to stable states).
Therefore, it is meaningful to find the optimal $\alpha$ for samples of different $S$ at which the reversal is fastest. For fixed $h_\text{mw} =0.035$ T, using the optimal  $f_0$ and  $R$ corresponding to  $S_0$, $S_1$, $S_2$ and $S_3$, we study the CCMP-driven magnetization reversal as a function of $\alpha$ for different $S$. FIG. \ref{fig:Fig4}(b) shows the dependence of switching time on the Gilbert damping for different $S$. For each $S$, there is certain value or range of $\alpha$ for which the switching time is minimal. For instance, the switching time is lowest at $\alpha=0.045$ for the sample of $S_1 = 10 \times 10 \text{\:nm}^2$. To be more clear, one may look at FIG. \ref{fig:Fig5}(a) and FIG. \ref{fig:Fig5}(b) which show the trajectories of magnetization reversal for $\alpha = 0.01$ and $\alpha = 0.045$ respectively and observed that for $\alpha = 0.045$, the reversal path is shorter. This is because, after crossing over the energy barrier, the larger damping dissipates the magnetization energy promptly and thus it leads to faster magnetization reversal. This finding suggests that larger $\alpha$ shows faster magnetization reversal.

\section{\label{sec:discussion}Discussions and Conclusions}

This study investigates the CCMP-driven magnetization reversal of a cubic sample at zero temperature limit and found that the CCMP with significantly smaller  $h_\text{mw}=0.035$ T, $f_0=18.8 \text{\:GHz}$  and  $R=0.32$ $ \text{ns}^{-1}$ than that of DCMP (i.e., $h_\text{mw}=0.045$ T, $f_0=21 \text{\:GHz}$  and $R$= 1.6 $ \text{ns}^{-1}$) can drive fast magnetization reversal. Since the frequency change of CCMP closely matches the magnetization precession frequency and thus it leads fast magnetization reversal with lower energy-cost.   Then we study the influence of demagnetization field$\slash$shape anisotropy field ${h}_\text{shape}$, on magnetization reversal process and the optimal parameters of CCMP. Interestingly we find that, with the increase of  $h_\text{shape}$, the parameters  $h_\text{mw}$ and $f_0$ of CCMP decreases with increasing $h_\text{shape}$.

So, we search a set of minimal parameters of CCMP for the fast ($t_{s}\sim$ 1 ns) magnetization reversal of the sample $22\times22\times 8$ nm$^3$ , and estimated as  $h_\text{mw}= 0.03 $ T, $f_0= 7.7 \text{\:GHz}$ and $R= 0.24$ $ \text{ns}^{-1}$ which are (significantly  smaller than that of DCMP) useful for device application. This is happened because with increase of ${h}_{\text{shape}}$, the effective anisotropy field decreases and thus the energy barrier (which separates two stable states) decreases. In addition, it is observed that the materials with the larger damping are better for fast magnetization reversal.
There is  a recent study  \cite{islam2020} reported that thermal effect assists the magnetization reversal i.e., thermal effect reduces  the controlling parameters of chirp microwave field pulse. Thus, it is expected that the parameters of CCMP also might be reduced further at room temperature. To generate such a cosine down-chirp microwave pulse, several recent technologies \cite{cai2013r,cai2010i} are available.
Therefore, the strategy of the cosine chirp microwave chirp driven magnetization reversal and other findings may lead  to realize the  fast and low-cost memory device.

\begin{acknowledgments}
This work was supported by the Ministry of Education (BANBEIS, Grant No. SD2019972). X. S. W. acknowledges the support from the Natural Science Foundation of China (NSFC) (Grant No. 11804045) and the Fundamental Research Funds for the Central Universities.
\end{acknowledgments}

\appendix

\section{Calculation of $\dot{\epsilon}$}
In this Appendix, we show the details of the derivation of $\dot{\epsilon}$ in Eq. \ref{eq:Io}. The rate of change of $\mathbf{h}_\text{mw}$ is
\begin{align*}
    \Dot{\mathbf{h}}_\text{mw} & = \frac{d\mathbf{h}_\text{mw}}{dt} \\
    & = \frac{d}{dt}\left(h_\text{mw} \left[ \cos\phi(t) \hat{\mathbf{x}} + \sin\phi(t) \hat{\mathbf{y}}\right] \right) \\
    & = h_\text{mw} \left[ -\sin\phi(t) \hat{\mathbf{x}} + \cos\phi(t) \hat{\mathbf{y}}\right] \frac{d\phi}{dt} \\
    & = h_\text{mw} \left[ -\sin\phi(t) \hat{\mathbf{x}} + \cos\phi(t) \hat{\mathbf{y}}\right] \left[ \frac{\phi(t)}{t} - \frac{d}{dt} \left(\frac{\phi(t)}{t}\right) t \right]
\end{align*}

The magnetization is given by
\begin{align*}
    \mathbf{m} & = m_x \hat{\mathbf{x}} + m_y \hat{\mathbf{y}} \\
    & = \sin\theta(t) \cos\phi_m(t) \hat{\mathbf{x}} + \sin\theta(t) \sin\phi_m(t) \hat{\mathbf{y}}
\end{align*}
where $\theta(t)$ is the polar angle and $\phi_m(t)$ is the azimuthal angle of the magnetization $\mathbf{m}$.

Substituting $m_x$ and $\Dot{\mathbf{h}}_\text{mw}$ in Eq. \ref{eng}, we get,
\begin{align*}
    \dot{\epsilon} & = - \mathbf{m} \cdot \Dot{\mathbf{h}}_\text{mw} \\
    & = h_\text{mw} \sin\theta(t) \left[ -\sin\phi(t) \cos\phi_m(t) + \cos\phi(t) \sin\phi_m(t)\right] \cdot \\
    & \qquad \qquad \qquad \qquad \qquad \qquad \qquad \left[ \frac{\phi(t)}{t} - \frac{d}{dt} \left(\frac{\phi(t)}{t}\right) t \right] \\
    & = h_\text{mw} \sin\theta(t) \sin \left( \phi(t) - \phi_m(t)\right) \left[ \frac{\phi(t)}{t} - \frac{d}{dt} \left(\frac{\phi(t)}{t}\right) t \right] \\
\end{align*}

Defining $\Phi (t) = \phi_m(t) - \phi(t)$, we have finally
\begin{align*}
    \dot{\epsilon} = h_\text{mw} \sin\theta(t) \sin \Phi(t) \left[ \frac{\phi(t)}{t} - \frac{d}{dt} \left(\frac{\phi(t)}{t}\right) t \right]
\end{align*}
where $\left[ \frac{\phi(t)}{t} - \frac{d}{dt} \left(\frac{\phi(t)}{t}\right) t \right]$ represents $\omega(t)$.

\nocite{*}

\section*{References}
\bibliography{aipsamp}

\end{document}